\documentclass[twocolumn,preprintnumbers,amsmath,amssymb]{revtex4}
\def\Journal#1#2#3#4{{#1} {\bf #2}, #3 (#4)}

\def\APP{Astropart. Phys.}

\def\NPB{Nucl. Phys. {\bf B}$\!$}
\def\NPBproc{Nucl. Phys. B (Proc. Suppl.)}
\def\PLB{Phys. Lett. B}

\def\PRL{Phys. Rev. Lett.}
\def\PRD{Phys. Rev. D}
\def\ZPC{Z. Phys. C}
\def\RMP{Rev. Mod. Phys.}
\def\MPLA{Mod. Phys. Lett {\bf A}$\!$}
\def\EPC{Eur. Phys. J. {\bf C}$\!$} 
\def\PREP{Phys. Rept.}
\def\ZPC{Zeitschrift f\"ur Physik {\bf C}$\!$}
\def\JHEP{JHEP}
\def\etal{{\sl et al.}} 
\usepackage{graphicx}% Include figure files

\begin{document}
\preprint{\vbox{\hbox{UCB-PTH/0023, LBNL-46431,MPI-PhT/2000-26}}}
\title{The Lower Bound on the Neutralino-Nucleon Cross Section}
\author{Vuk Mandic}
\affiliation{Department of Physics, 
University of California, 
Berkeley, CA~~94720, USA}
\author{Aaron Pierce}
\author{Hitoshi Murayama}
\affiliation{Department of Physics, 
University of California, 
Berkeley, CA~~94720, USA;\\ 
Theory Group, 
Lawrence Berkeley National Laboratory, 
Berkeley, CA~~94720, USA}
\author{Paolo Gondolo}
\affiliation{ Max-Planck-Institut f\"{u}r Physik, F\"{o}hringer Ring 6, D-80805
  M\"{u}nchen, Germany}

\date{\today}

\begin{abstract}
We examine if there is a lower bound on the detection cross section, 
$\sigma_{\chi-p}$, for the neutralino dark matter in the MSSM. If we 
impose the minimal supergravity boundary conditions as well as the 
``naturalness'' condition, in particular $m_{1/2}<300$~GeV, we show 
that there is a lower bound of $\sigma_{\chi-p}>10^{-46}$~cm$^{2}$.  
We also clarify the origin for the lower bound.  Relaxing either of 
the assumptions, however, can lead to much smaller cross sections.
\end{abstract}

\maketitle

\setcounter{footnote}{0}
\setcounter{page}{1}
\setcounter{section}{0}
\setcounter{subsection}{0}
\setcounter{subsubsection}{0}

%%%%%%%%%%%%%%%%%%%%%%%%%%%%%%%%%%%%%%%%%%%%%%%%%%%%%%%%%%%%%%%%%%%%%%%

%\vskip 0.3in

\section{Introduction}
Supersymmetry is considered to be a compelling extension to the 
Standard Model for several reasons.  For example, it stabilizes scalar 
masses against radiative corrections, allowing theories with 
fundamental scalars to become natural.  For a review see 
\cite{HMSusy}.  Supersymmetry also weighs in on the dark matter 
problem: stars and other luminous matter contribute a small fraction 
of the critical density, $\Omega_{lum} = (0.003 \pm 0.001) h^{-1}$, 
while the amount of matter known to exist from its gravitational 
effects (both at galaxy and cluster of galaxies scales) is much 
larger, $\Omega_{M} = 0.35 \pm 0.07$ (Ref.~\cite{tysonturner}).  
Furthermore, most of the missing matter seems to be non-baryonic in 
nature.  The experimental motivation behind the dark matter problem 
and different search strategies are discussed in more detail in 
\cite{sadoulet,JKG,KolbTurner}.

In supersymmetric models, R-parity is often imposed to avoid 
weak-scale proton decay or lepton number violation.  Imposing this 
symmetry also yields an ideal fermionic dark matter candidate.  
Namely, in supersymmetric models with R-parity, the lightest 
supersymmetric particle (LSP) is stable and it could conceivably make 
up a substantial part of the dark matter in the galactic halo.

Here we investigate the direct detection of such a particle.  There 
have been many such studies in the literature 
\cite{JKG,crosssections,nnsreexamined,bottino,GGR,bergstromgondolo}.  
More recently, there have been discussions of numerous variables that 
can effect direct detection.  These studies include an investigation 
of the effect of the rotation of the galactic halo \cite{halo}, the 
effects of the uncertainty of the quark densities within the nuclei 
\cite{bottino2,NathCorsetti,quarkdensity}, possible CP violation 
\cite{Chatto,CP}, and non-universality of gaugino masses 
\cite{NathCorsetti,NathCorsetti2}.  Here we attempt to address the following
question: Is there a minimum cross section for the elastic 
scattering of neutralinos off of ordinary matter?  Naively, it would 
seem that a judicious choice of parameters might allow a complete 
cancellation between different diagrams.  After all, the parameter 
space is very large in the general Minimal Supersymmetric 
Standard Model (MSSM), and even for a very restrictive framework such 
as the minimal supergravity (mSUGRA), the number of parameters is 
still quite large.  We will show that there nonetheless exists a 
minimum cross section in the mSUGRA framework. However, we will also 
show that this result strongly depends on the assumptions of the 
framework, such as unification of different parameters at the GUT 
scale, radiative electroweak symmetry breaking and naturalness.

This argument is of great importance when considering the upcoming 
direct 
detection experiments. For the mSUGRA framework, one expects that the 
future 
ambitious direct detection experiments can explore most of the 
parameter
space. However, we find that the detection picture is not quite as 
rosy for 
a more general MSSM framework.

\section{Definitions and Approach}
We adopt the following notation for the superpotential and soft 
supersymmetry
breaking potential in the MSSM:
\begin{eqnarray}
W & = & \epsilon_{ij} (-\hat{e}_R^* h_E \hat{l}_L^i \hat{H}_1^j
	-\hat{d}_R^* h_D \hat{q}_L^i \hat{H}_1^j \nonumber \\
& & + \hat{u}_R^* h_U \hat{q}_L^i \hat{H}_2^j - \mu \hat{H}_1^i 
\hat{H}_2^j), \\
V_{soft} & = & \epsilon_{ij} (\widetilde{e}_R^* A_E h_E 
\widetilde{l}_L^i H_1^j
	+ \widetilde{d}_R^* A_D h_D \widetilde{q}_L^i H_1^j \nonumber \\
& & - \widetilde{u}_R^* A_U h_U \widetilde{q}_L^i H_2^j
	- B \mu \hat{H}_1^i \hat{H}_2^j +h.c.) \nonumber \\
& & + H_1^{i*} m_{H_1}^2 H_1^i + H_2^{i*} m_{H_2}^2 H_2^i \nonumber \\
& & + \widetilde{q}_L^{i*} M_Q^2 \widetilde{q}_L^i
	+ \widetilde{l}_L^{i*} M_L^2 \widetilde{l}_L^i
	+ \widetilde{u}_R^* M_U^2 \widetilde{u}_R \nonumber \\
& & + \widetilde{d}_R^* M_D^2 \widetilde{d}_R
	+ \widetilde{e}_R^* M_E^2 \widetilde{e}_R \nonumber \\
& & + \frac{1}{2}(M_1 \widetilde{B} \widetilde{B} 
	+ M_2 \widetilde{W}^a \widetilde{W}^a
	+ M_3 \widetilde{g}^b \widetilde{g}^b).
\end{eqnarray}
Here the $h$'s are Yukawa 
couplings, the $A$'s are trilinear couplings, the $M_{Q,U,D,L,E}$ are 
the squark and slepton mass parameters, the $M_{1,2,3}$ are gaugino 
mass parameters and $m_{H_1}$, $m_{H_2}$, $\mu$, and $B$ are Higgs 
mass parameters.  The $i$ and $j$ are $SU(2)_L$ indices, and are made 
explicit, so as to make our sign conventions clear.  $SU(3)$ indices 
are suppressed.  In the R-parity invariant MSSM the LSP is usually a 
neutralino - a mixture of bino, neutral wino and two neutral higgsinos.  
In our notation, the neutralino mass matrix reads
\begin{equation}
\small{
\left( \begin{array}{cccc}
M_{1} & 0 & -m_{Z} s_{{\theta}_{W}} c_{\beta} &
+m_{Z} s_{\theta_{W}} s_{\beta} \\
0 & M_{2} & +m_{Z} c_{\theta_{W}} c_{\beta} &
-m_{Z} c_{\theta_{W}} s_{\beta} \\
-m_{Z} s_{\theta_{W}} c_{\beta} & +m_{Z} c_{\theta_{W}} c_{\beta} &
0 & -\mu \\
-m_{Z} s_{\theta_{W}} s_{\beta} & -m_{Z} c_{\theta_{W}} s_{\beta} & 
-\mu & 0
\end{array} \right)}
\label{massmat}
\end{equation}
Here $s_{\beta}=\sin\beta$, $c_{\beta}=\cos\beta$, 
$s_{\theta_{W}}=\sin\theta_{W}$, and $c_{\theta_{W}}=\cos\theta_{W}$.  
The physical states are obtained by diagonalizing this matrix.  The 
lightest neutralino can be written in the form:
\begin{equation}
\chi^{0}_{1}=N_{11} \tilde{B} + N_{12} \tilde{W}_{3} + N_{13} 
\tilde{H}_{1}^{0} +N_{14} 
\tilde{H}_{2}^{0}.
\label{neutralinocontent}
\end{equation}

\begin{figure}[b]
\includegraphics[width=\columnwidth]{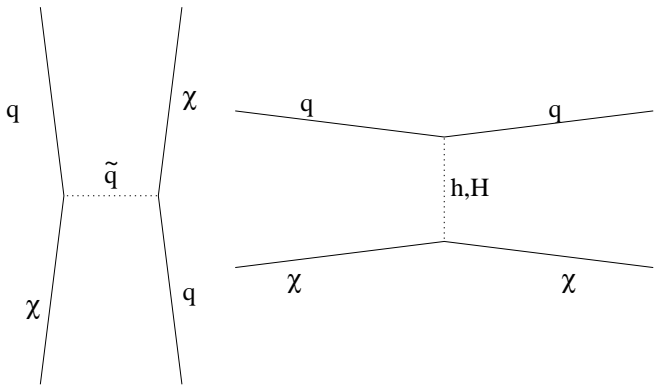}
\label{fig:diagrams}
\caption{The leading diagrams for direct detection.  Note that there 
is also a $u$-channel diagram for squark exchange.  There are also 
diagrams where the neutralino scatters off of gluons in the nucleon 
through heavy squark loops.}
\label{diagrams}
\end{figure}

We are interested in spin independent scattering of neutralinos off of 
ordinary matter.  This contribution dominates in the case of detectors 
with large nuclei, such as Ge \cite{Goodman}.  As discussed in the 
literature, in most situations the dominant contribution to the spin 
independent amplitude is the exchange of the two neutral Higgs bosons, 
although in some cases the contribution of the squark exchange and 
loop corrections are substantial.  The relevant tree-level diagrams 
are shown in Figure~\ref{diagrams}.

We use the DarkSUSY package \cite{DarkSUSY} in this study.  
This code has the following inputs: $M_{1,2,3}$, 
$\mu$, the ratio of the vevs of the two Higgs bosons ($\tan \beta = 
v_2 / v_1$), the mass of the axial Higgs boson ($m_A$), the soft 
masses of the sparticles ($M_{Q,U,D,L,E}$) and the diagonal components 
of the trilinear coupling matrices ($A_{E,D,U}$).  All inputs are to 
be supplied at the weak scale.  DarkSUSY then calculates the particle 
spectrum, widths and couplings based on the input parameters.  It 
evaluates the cross section for scattering of neutralinos off protons 
and neutrons, following Ref.~\cite{bergstromgondolo}.  It also 
evaluates the relic density of the neutralinos for the given input 
parameters following \cite{relicdensity}, which includes the 
relativistic Boltzmann averaging, sub-threshold and resonant 
annihilation and coannihilation processes with charginos and 
neutralinos.  Furthermore, DarkSUSY checks for the current constraints 
obtained by experiments, including the $b \rightarrow s + \gamma$ 
constraint \cite{implementbsg,DSbounds}.

\section{mSUGRA Framework}
\subsection{Definition of the Framework}
In this section, we briefly outline the mSUGRA framework.  
In mSUGRA, one makes several assumptions:

\begin{itemize}
\item There exists a Grand Unified Theory (GUT) at some high energy 
scale.  Consequently, the gauge couplings unify at the GUT scale.  The 
value of the couplings at the weak scale determines the GUT scale to 
be $\approx 2 \times 10^{16}$ GeV. The gaugino mass parameters also 
unify to $m_{1/2}$ at the GUT scale.

\item Other unification assumptions are: the scalar mass parameters 
unify to a value denoted by $m_{0}$ and the trilinear couplings unify 
to $A_0$ at the GUT scale.  Using the MSSM renormalization group 
equations (RGEs) we evaluate all parameters at the weak scale.  We 
choose to do this using the one loop RGEs that can be found, for 
example, in \cite{RGEs} or \cite{SUGRAWorking}.

\item Radiative electroweak symmetry breaking (REWSB) is imposed: 
minimization of the one-loop Higgs effective potential at the 
appropriate scale fixes $\mu^2$ and $m_A$ (we follow the methods of 
Refs.~\cite{ArnowittNathEWSB} and \cite{effpot}).  For completeness, 
we reproduce the equation for $\mu^2$ at tree level:
\begin{equation}
\mu^2=\frac{m_{H_{1}}^2-m_{H_{2}}^2 \tan^2 \beta}{\tan^2 \beta -1} 
-\frac{1}{2}M_{Z}^2 .
\label{eqn:EWSB}
\end{equation}
\end{itemize}

With these assumptions, the mSUGRA framework allows four free 
parameters ($m_0$, $m_{1/2}$, $A_0$ and $\tan\beta$).  Also, the sign 
of $\mu$ remains undetermined.  Starting with these parameters we 
determine all of the input parameters for the DarkSUSY code.  We allow 
the free parameters to vary in the intervals
\begin{eqnarray}
0 < m_{1/2} < 300 \, {\rm GeV}, \; 95< m_{0} < 1000 \, {\rm GeV}, 
\nonumber \\
-3000 < A_{0} < 3000 \, {\rm GeV}, \; 1.8 < \tan \beta< 25. 
\label{paramspace}
\end{eqnarray}
The upper bounds on these parameters come from the naturalness 
assumption: one of the reasons for using supersymmetry is its ability 
to naturally relate high and low energy scales; as a result, no 
parameter in the theory should be very large. Moreover, the upper bound on
$m_{1/2}$ makes our data set insensitive to the stau-neutralino
coannihilations, which are not included in the calculation of relic density 
performed by DarkSUSY. We have explicitly checked that there are no
models in our data set which would be cosmologically allowed {\bf only}
if the stau coannihilations 
were included. We will also see later that expanding the upper bound on
$m_{1/2}$ has serious consequences regarding the lower bound on the elastic 
scattering cross section. Also, the low value of 
$\tan\beta$ is set by the requirement that the top Yukawa coupling 
does not blow up before the GUT scale is reached.

Before we present the detailed analysis of cross section, a few 
remarks are in order.  First, the $b\rightarrow s + \gamma$ constraint 
eliminates large portions of the $\mu < 0$ parameter space, in 
agreement with \cite{bsgamma}, \cite{nohiggsino}.  Second, for both 
$\mu>0$ and $\mu<0$, we find no higgsino-like LSP models that are 
cosmologically important, in agreement with \cite{nohiggsino}.

\begin{figure}[hbtp]
\includegraphics[width=\columnwidth]{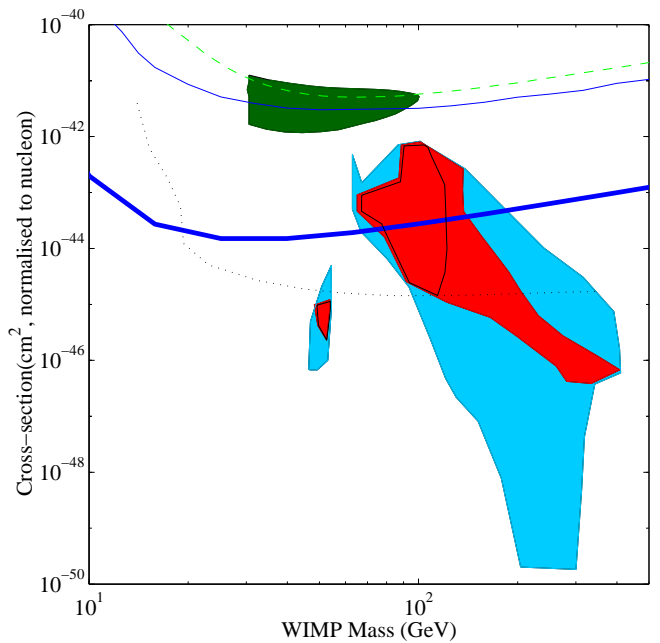}
\caption{The cross section for spin-independent $\chi$-proton 
scattering is shown.  Current accelerator bounds, 
including $b \rightarrow s + \gamma$ are imposed through the DarkSUSY 
code.  The top dark region is the DAMA allowed region at 3$\sigma$ CL. 
The dashed curve is the DAMA $90\%$ CL exclusion limit from 1996 
(obtained using pulse-shape analysis).  The thinner solid curve is the current 
CDMS $90\%$ CL exclusion limit, the thicker solid curve is the projected
exclusion limit for CDMS II experiment and the dotted curve is the projected 
exclusion limit for the GENIUS experiment.  The light shaded regions 
denote mSUGRA models passing the $0.025 < \Omega h^2 < 1$ constraint and
with the upper bound $m_{1/2}<1$ TeV. As discussed in the text, this region
shows that including stau coannihilations into the relic density calculation
has a dramatic effect on the lower bound on $\sigma_{\chi-p}$ for 
$m_{1/2}>300$ GeV ($M_{\chi}>120$ GeV). Restricting the relic density further
to $0.052<\Omega h^2<0.236$ yields the darker regions within the lighter 
regions and restricting also the upper bound on $m_{1/2}<300$ GeV (Eq. 
(\ref{paramspace})) gives the regions bounded by the closed solid curves.}
\label{csvsmchim3}
\end{figure}

We plot the variation of the spin independent cross section versus the 
neutralino mass in Figure~\ref{csvsmchim3}.  Note that the complete 
allowed region is split into two parts by the annihilation channel into 
$W^+ W^-$, which affects the relic density of neutralino. We consider two
relic density constraints. Since the present observations favor the Hubble 
constant $h = 0.7 \pm 0.1$ and the total matter density 
$\Omega_M = 0.3 \pm 0.1$, of which baryons contribute 
$\Omega_b h^2 \approx 0.02$, we consider the range $0.052 < \Omega_{\chi} h^2
< 0.236$. However, we also examine effects of relaxing the relic density
constraint to $0.025 < \Omega_{\chi} h^2 < 1$. 
%In particular, since we do not include the neutralino-stau coannihilations
%in the relic density calculation, this relaxed relic density cut depicts 
%the effect that inclusion of the stau coannihilations could have on the 
%lower bound on $\sigma_{\chi-p}$. 
The upper bound on 
$\sigma_{\chi-p}$ (of the theoretically allowed regions) comes from the lower 
bound on the relic density. The lower bound on $M_{\chi}$ comes from 
the existing constraints from the accelerator experiments.  The upper bound on 
$M_{\chi}$ is a combination of the upper bound on relic density
and of the bounds on the free parameters.  
The lower bound on $\sigma_{\chi-p}$ is not yet well understood, and it is the 
subject of this paper.  
%Notice that a similar plot has already 
%appeared - Fig.~1 in Ref.~\cite{NathCorsetti}.  We find good 
%agreement with this reference.  
Figure \ref{csvsmchim3} also includes 
some recent and future direct detection experimental results 
\cite{DAMA1,DAMA2,CDMS,CDMSII,GENIUS}.  Note that the parameter space defined 
by Eq.~(\ref{paramspace}) corresponds to the region of $\sigma_{\chi 
- p}$ - $M_{\chi}$ plane bounded by the two closed solid lines. 
Therefore, $\sigma_{\chi-p}>10^{-46} {\rm cm^2}$ for these models, or 
equivalently, assuming a $^{73}{\rm Ge}$ target, the dark matter 
density $\rho_D = 0.3 \; {\rm GeV} c^{-2} {\rm cm^{-3}}$, the WIMP 
characteristic velocity $v_0 = 230 \; {\rm km \; s^{-1}}$ and 
following Ref.~\cite{smithlewin}, the event rate $R>0.1 \; {\rm 
ton^{-1} day^{-1}}$.  Hence, the most ambitious future direct 
detection experiments may be able to explore a large portion, if not 
all, of the these models.

\subsection{Results and Analysis}
\begin{figure}[hbtp]
\includegraphics[width=\columnwidth]{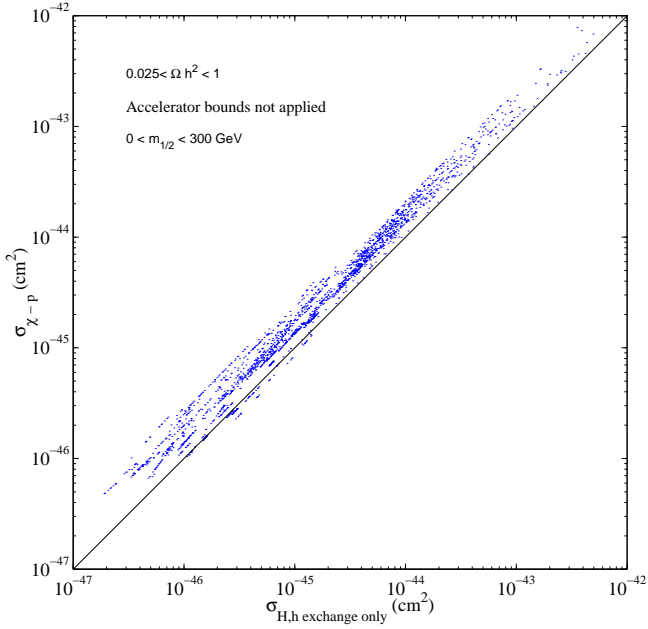}
\label{fig:cs_vs_sigmahH}
\caption{The cross section for spin-independent $\chi$-proton 
scattering: the complete (DarkSUSY) calculation is shown on the $y$ 
axis and the contribution to the cross section from the Higgs bosons 
exchange alone is shown on the $x$ axis.  A relatively conservative 
relic density cut is applied, $0.025<\Omega h^{2}<1$.}
\label{csvssigmaHh}
\end{figure}

As mentioned above, the dominant contribution to spin independent 
elastic scattering is usually the Higgs boson exchange.  Figure 
\ref{csvssigmaHh} illustrates this relationship within our results.  
The nearly perfect 45 degree line in the figure indicates good 
agreement between the total cross section as evaluated by DarkSUSY and 
the cross section calculated including the exchange of Higgs bosons 
only (in the approximation explained below).  We will, therefore, 
concentrate on the Higgs boson exchange and will postpone the 
discussion of squark exchange to the end of this section.

The contribution of Higgs boson exchange can be found in the 
literature 
\cite{JKG,bergstromgondolo,NathCorsetti,Chatto,ArnowittNathExpansion,%
griest,barbieri}.  It is of the following form:
\begin{equation}
\sigma_{h,H} \sim |(f_u + f_c + f_t) A^u + (f_d + f_s + f_b) A^d |^2,
\label{sigma}
\end{equation}
where $f_u \approx 0.023, f_d \approx 0.034, f_s \approx 0.14, 
f_c = f_t = f_b \approx 0.0595$ 
parametrize the quark-nucleon matrix elements and
\begin{eqnarray}
A^u &=& {g_2^2\over 4M_W} \Bigl({F_h\over m_h^2}
 {\cos\alpha_H\over \sin\beta} + {F_H\over m_H^2} {\sin\alpha_H\over 
\sin\beta}\Bigr),
\label{aueq}\\
A^d &=& {g_2^2\over 4M_W} \Bigl(-{F_h\over m_h^2} {\sin\alpha_H\over 
\cos\beta} 
 + {F_H\over m_H^2} {\cos\alpha_H\over \cos\beta}\Bigr),
\label{adeq}
\end{eqnarray}    
and 
\begin{eqnarray}
F_h &=&(N_{12}-N_{11} \tan\theta_W)(N_{14} \cos\alpha_H+N_{13} 
\sin\alpha_H) \nonumber \\
F_H &=& (N_{12}-N_{11} \tan\theta_W)(N_{14} \sin\alpha_H-N_{13} 
\cos\alpha_H). 
\label{fhH}
\end{eqnarray}
The $N$'s are the coefficients appearing in 
Eq.~(\ref{neutralinocontent}) and $\alpha_H$ is the Higgs boson mixing 
angle (defined after radiative corrections have been included in the Higgs 
mass matrix).  Note that there is an upper bound on 
the light Higgs boson mass in the MSSM, given by $m_h < 130$ GeV 
\cite{higgsmass}.  $A^u$ 
represents the amplitude for scattering off an up-type quark in a 
nucleon, while $A^d$ represents the amplitude for scattering off a 
down-type quark in the nucleon.  
Since all models we generate have a bino-like 
neutralino, $\mu>M_1$ and $\mu > M_Z$.  Then, following 
Ref.~\cite{ArnowittNathExpansion}, we can expand the $N_{1i}$'s out in 
powers of $\frac{M_{Z}}{\mu}$.  We reproduce their result here:
\begin{eqnarray}
\label{expansion}
N_{11} & \approx &1, \\\
N_{12} & \approx &
-{1\over2}{M_{Z}\over\mu}{\sin2\theta_W\over{(1-M_{1}^{2}/\mu^{2})}}{M_{Z}
\over{M_{2}}-M_{1}}\left[\sin2\beta +
{M_{1}\over\mu}\right], \nonumber \\
N_{13} & \approx &{M_{Z}\over\mu}{1\over{1-M_{1}^{2}/\mu^{2}}}
\sin\theta_{W}\sin\beta\left[1+{M_{1}\over\mu} \cot\beta\right] ,
\nonumber \\
N_{14} & \approx &-{M_{Z}\over\mu}{1\over{1-M_{1}^{2}/\mu^{2}}}
\sin\theta_{W} \cos\beta\left[1+{M_{1}\over\mu} \tan\beta\right] 
\nonumber.
\end{eqnarray} 

\begin{figure}[hbtp]
\includegraphics[width=\columnwidth]{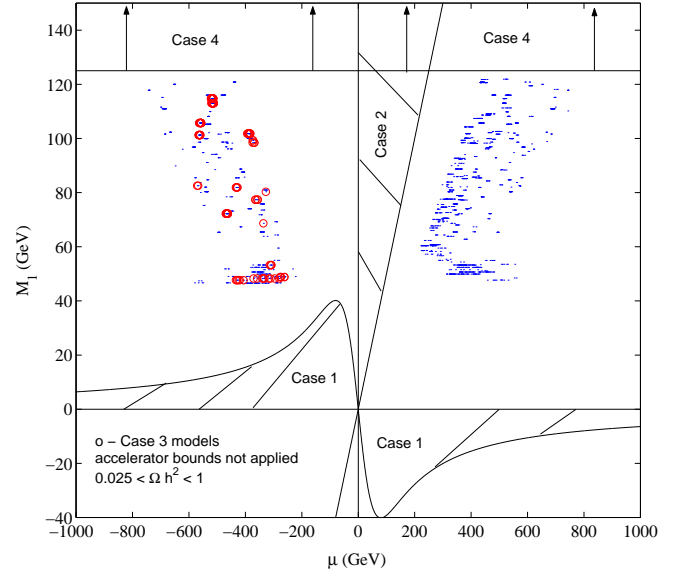}
\caption{This figure illustrates 
the regions of $M_1 - \mu$ plane (as defined by the Eq. (\ref{paramspace})) 
in which the different 
cases discussed in the text can be satisfied. For the models shown, 
the accelerator constraints 
were not applied, but a relatively conservative relic density cut 
($0.025<\Omega h^{2}<1$) was applied. Note that almost all of 
the models of Case 3 shown in this plot are ruled out by the 
accelerator constraints. Also, there are no models satisfying the Case 5 for
the parameter space defined by Eq. (\ref{paramspace}).}
\label{m1vsmu}
\end{figure}

In this approximation, we can identify five possible situations in which
the neutralino-proton elastic scattering cross section is relatively small.
Figure~\ref{m1vsmu} illustrates which 
regions of the $M_1$-$\mu$ plane satisfy the conditions of the five 
cases.  In the following, we will use the approximation of 
Eq.~(\ref{expansion}).

\begin{itemize}
\item Case 1: $N_{12} - N_{11} \tan\theta_W = 0$ would make both $F_h$ 
and $F_H$ vanish.  Intuitively, this is reasonable since this 
condition implies that the neutralino is a pure photino, and the tree 
level Higgs coupling to the photino vanishes.  Using 
Eq.~(\ref{expansion}), we may rewrite this condition as
\begin{equation}  
\mu(M_2-M_1) = -M_Z^2\cos^2\theta_W 
\left(\sin2\beta+\frac{M_1}{\mu}\right).
\end{equation} 
Since $M_2>M_1>0$, the last equation can be satisfied only if $\mu<0$.  
If we use the GUT relationship betweeen $M_1$ and $M_2$, we get
\begin{equation}
\mu M_1 \left(\frac{3\cos^2 \theta_W}{5\sin^2\theta_W}-1\right) 
= -M_Z^2 \cos^2\theta_W \left(\sin2\beta+\frac{M_1}{\mu}\right)
\end{equation} 
or, equivalently, for $\mu$ and $M_1$ in units of GeV
\begin{equation}
M_1 = \frac{-6449\sin2\beta}{\mu+\frac{6449}{\mu}}.
\label{condition1}
\end{equation}
As $\sin2\beta$ ranges from 0 to 1, Eq.~(\ref{condition1}) spans the 
dashed regions marked 'Case 1' in  Figure \ref{m1vsmu}.  
For $\sin2\beta = 1$, Eq.~(\ref{condition1}) implies $M_{\chi} \approx 
M_1<40$ GeV. This can be read directly from Figure~\ref{m1vsmu}.  Note 
that such low values of $M_{\chi}$ are excluded by the relic density 
constraint and the current experimental limits (as shown in 
Figure~\ref{csvsmchim3} and Ref.~\cite{NathCorsetti}).  The constraint 
on $M_{\chi}$ is even stronger for other values of $\sin2\beta$, so we 
conclude that this condition cannot be satisfied in the mSUGRA 
framework.

\item Case 2: $N_{14} \sin\alpha_H - N_{13} \cos\alpha_H = 0$ would 
make only $F_H$ vanish.  In our approximation, this condition 
translates into
\begin{equation}
M_1/\mu = -\cot\alpha_H - \cot\beta. 
\label{fHcondition}
\end{equation}
To understand the meaning of this condition better, we will use the 
tree level relationship between $\alpha_{H}$ and $\beta$:
\begin{eqnarray}
\cot 2\alpha_H = k \cot2\beta, \; k=\frac{m_A^2 - M_Z^2}{m_A^2 + 
M_Z^2},
\label{cot2alpha}
\end{eqnarray}
which, after some trigonometric manipulations, yields
\begin{eqnarray}
\cot \alpha_H & = & \frac{k}{2} \left(\frac{1}{\tan\beta} - 
\tan\beta\right)  
\nonumber \\
&  &  - \sqrt{\frac{k^2}{4}\Bigl(\frac{1}{\tan^2\beta} + \tan^2\beta 
-2\Bigr) +1}.
\label{cotalpha}
\end{eqnarray}

Since both terms on the right hand side of Eq.~(\ref{cotalpha}) are 
negative because $\tan\beta>1$, the minimum of $|\cot\alpha_H| = 1$ occurs 
when $k=0$ (or, equivalently, when $m_A = M_Z$).
Then, since $\tan\beta>1.8$, the condition of Eq.~(\ref{fHcondition}) 
becomes 
\begin{equation}
\label{case2numeric}
\frac{M_1}{\mu} > 0.5. 
\end{equation}
At this point, we would like to formulate a relationship between 
$\mu$ and $M_1$ resulting from the RGEs and REWSB assumptions. 
In Ref.~\cite{SUGRAWorking}, an approximate solution (based on the 
expansion 
around
the infrared fixed point) to the RGEs for the Higgs mass 
parameters, $m_{H_1}$ and $m_{H_2}$, is presented. Assuming that the 
value of 
the top Yukawa coupling is relatively close to the infrared fixed 
point, 
we can write:
\begin{eqnarray}
m_{H_{1}}^2 \approx m_0^2+0.5 m_{1/2}^{2}, \nonumber \\
m_{H_{2}}^2 \approx -0.5 m_0^2 - 3.5 m_{1/2}^{2}.
\end{eqnarray}

These equations, coupled with Eq.~(\ref{eqn:EWSB}), yield a value for 
$\mu^2$ in terms of $m_{1/2}, m_{0}$, and $\tan \beta$. Using the GUT 
relationship:
\begin{equation}
M_1 = m_{1/2} \frac{\alpha_{1}(m_{Z})}{\alpha_{GUT}},
\label{m1mhalf}
\end{equation}
we get a roughly linear relationship between $\mu$ and $M_1$:
\begin{equation}
M_1 = (0.3|\mu| - 60) \pm 40.
\end{equation}
The spread $\pm 40$ comes from the variation in $m_0$ and $\tan\beta$ 
and the linearity breaks down somewhat at the low values of $M_1$.  
The empirical relationship that we obtain from running the code (see Figure ~\ref{m1vsmu}) is 
very similar:
\begin{equation}
M_1 = (0.3|\mu| - 40) \pm 25.
\label{expm1mu}
\end{equation}
The smaller spread comes from the application of the relic density cut 
and the current experimental limits.  In any case, we conclude that 
$M_1/\mu>0.3$ is not allowed in the mSUGRA framework.  This is in 
conflict with Eq.~(\ref{case2numeric}), implying that the Case 2 
cannot be satisfied.  Note that the tree level relationship in 
Eq.~(\ref{cot2alpha}) is altered at higher orders, but we 
have checked that this does not affect the final conclusion.

\item Case 3: $N_{14} \cos\alpha_H + N_{13} \sin\alpha_H = 0$ would 
make $F_h$ vanish.  Manipulation of this condition using 
Eq.~(\ref{expansion}) yields
\begin{equation}
M_1/\mu = \tan\alpha_H - \cot\beta. 
\label{fhvanish}
\end{equation}

Figure \ref{csvssigmaH} shows that when this condition is 
(approximately) satisfied, the contribution of
the heavy Higgs boson exchange to the elastic scattering cross section is the
largest.  Hence, even if this 
condition is satisfied, $\sigma_{\chi-p}$ cannot be arbitrarily small due to 
heavy Higgs boson exchange (along with other channels such as the squark 
exchange).  Of course, this assumes that there is some 
upper bound on the heavy Higgs mass, which is true 
simply because the parameter space is bounded. Note, also, that since 
$\tan\alpha_H<0$, the condition for the vanishing of the light Higgs 
boson contribution can be satisfied only for $\mu<0$. It is interesting, 
however, that most of the models in our data set 
that approximately satisfy this condition are excluded by the bounds on 
Higgs boson masses.

\begin{figure}[hbtp]
\includegraphics[width=\columnwidth]{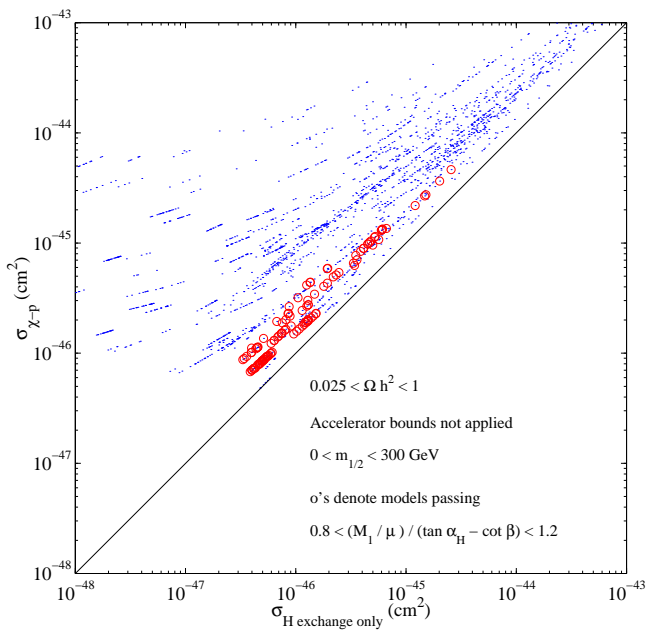}
\label{fig:cs_vs_sigmaH}
\caption{The proton-neutralino scattering cross section for models 
where the condition in Case 3 is approximately satisfied (in particular, 
models satisfying $0.8 < \frac{M_1/\mu}{\tan\alpha_H - \cot\beta} < 1/2$ are
denoted by o's). 
The complete (DarkSUSY) calculation is shown on the $y$-axis and the 
contribution to the cross section due only to the exchange of the 
heavy Higgs boson, is shown on the $x$-axis.  A relatively 
conservative relic density cut is applied, $0.025<\Omega h^{2}<1$.
Most of the models denoted by o's in this plot are excluded by the 
accelerator bounds on the Higgs boson masses.}
\label{csvssigmaH}
\end{figure}

\item Case 4: There is one more way of making both $F_h$ and $F_H$ small, 
and that is by making $\mu$ very large ($N_{12}$, $N_{13}$, and 
$N_{14}$ all contain $\mu$ in denominator).  However, this possibility 
is limited by naturalness assumption: $\mu$ is kept below $\approx 
850$ GeV by the upper bound we have chosen on $m_{1/2}$.  Hence, in 
mSUGRA framework, the naturalness assumption also keeps the cross 
section from vanishing.

\item Case 5: We now consider the possibility of complete cancellation of
terms in the Equation (\ref{sigma}).
Let us examine the relative signs of $F_h$, $F_H$ and 
$\mu$.  Since $\tan\beta>1.8$ and $M_1/\mu\approx0.3$, it follows from 
Eq.~(\ref{expansion}) that $|N_{13}|>|N_{14}|$.  Then, since 
$\cot\alpha_H<-1$ (and $\sin\alpha_H<0$), the $N_{13}$ term dominates 
over the $N_{14}$ term in $F_H$ (Eq.~(\ref{fhH})).  Hence, $F_H/\mu>0$ 
always, consistent with the analysis of Case 2 above.  The situation 
is somewhat more complicated in the case of $F_h$.  For $\mu>0$, following a 
similar analysis we get $F_h/\mu>0$.  Then, the interference between the 
two terms in $A^u$ (Eq.~(\ref{aueq})) is destructive and the 
interference between the two terms in $A^d$ (Eq.~(\ref{adeq})) is 
constructive.  Furthermore, in Eq.~(\ref{sigma}) we see that $A^u$ 
gets multiplied by a much smaller form factor than $A^d$.  As a 
result, $A^d$ strongly dominates over $A^u$ in Eq.~(\ref{sigma}), 
preventing $\sigma_{\chi-p}$ from vanishing.

On the other hand, if $\mu<0$ and if $\mu > - M_1 \tan\beta$, 
$N_{14}$ can change sign relative to $\mu$. This change of sign can 
propagate through Equations (\ref{fhH}), (\ref{adeq}), and (\ref{sigma}),
so that the $A^u$ and $A^d$ terms in the Eq. (\ref{sigma}) are of
opposite sign. If the parameters are tuned properly, a complete 
cancellation of these terms can be obtained. However, this
cancellation happens only if $M_{\chi} (\approx M_1) > 120$ GeV, which
is not allowed due to the upper bound on $m_{1/2}$. We conclude that in the
parameter space defined by Eq.~(\ref{paramspace}), $A^d$ always 
dominates over $A^u$, so $\sigma_{\chi-p}$ does not vanish.

If the upper bound on $m_{1/2}$ is relaxed to 1 TeV, the situation changes 
significantly. The complete cancellation discussed above is now
allowed to happen. As shown in Figure \ref{fdadfuau}, the lowest
values of $\sigma_{\chi-p}$ are reached exactly when the $A^d$ and $A^u$
terms cancel each other. As shown in Figure \ref{csvsmchim3}, the 
stringent relic rensity cut $0.052 < \Omega h^2 < 0.236$ 
rules out all of these 
models. However, we do not include the stau-neutralino coannihilations in
the calculation of the relic density, so this result should be taken with
caution (see, for example, \cite{ellisstau,arnowittstau}). 
In fact, our data set contains models in which the stau-neutralino
coannihilation could make a difference. For this reason, the Figure
\ref{csvsmchim3} also contains the allowed region for a conservative 
relic density cut $0.025< \Omega h^2 < 1$. Clearly, this cut allows
models with very low $\sigma_{\chi-p}$ values.
\end{itemize}

\begin{figure}[hbtp]
\includegraphics[width=\columnwidth]{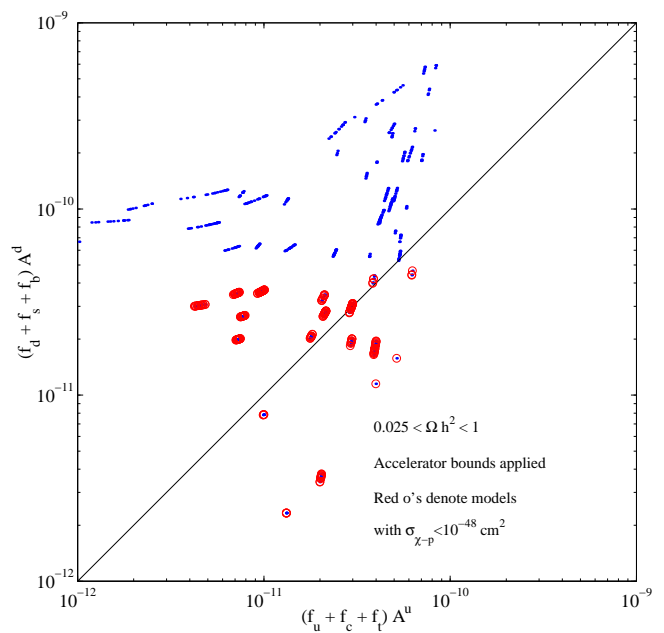}
\label{fig:fdad_vs_fuau}
\caption{The absolute values of the $A^d$ and the $A^u$ terms of Eq. 
(\ref{sigma}) 
are shown. Only $A^d < 0$ (for which $\mu<0$) are shown, satisfying a 
conservative relic density constraint $0.025 < \Omega h^2 < 1$, 
passing the accelerator bounds, and with $m_{1/2} < 1$ TeV. The o's denote
the lowest $\sigma_{\chi-p}$ models ($<10^{-48} \; {\rm cm^2}$). 
Clearly, the lowest values of 
$\sigma_{\chi-p}$ are achieved when the $A^d$ and $A^u$ terms cancel 
each other out.}
\label{fdadfuau}
\end{figure}

With the above discussion in hand, let us go back and consider the 
squark exchange.  The complete calculation of the squark exchange 
contribution is fairly complex.  However, good insights can be gained 
by making several simplifying assumptions.  First of all, the 
contribution of the squark exchange can be roughly approximated by the 
contribution of the exchange of the $u$, $d$, and $s$ squarks.  In this case, 
the contribution of the squark exchange to the cross-section can be 
written as
\begin{equation}
\sigma_{\tilde{q}} \sim |f_u B_u + f_d B_d + f_s B_s |^2,
\end{equation} 
where $f_i$ is as defined above, and the $B_j$ represent the amplitude 
for scattering off of a quark of type $j$ in the nucleon.  Furthermore, 
in the following considerations we neglect the left-right mixing in 
these light squarks.  This should be true over a large class of 
models, as the off-diagonal elements in the squark mass matrix are 
proportional to the corresponding quark mass.  Let us also neglect the 
mass splitting of the two different squarks.  Also, since $f_s \gg
f_u$, $f_d$, we can neglect all but the $B_s$ term.  In this 
approximation, following Ref.~\cite{Chatto}, we can write:
\begin{equation}
B_s=-\frac{1}{4 m_s} \frac{1}{M_{\tilde{s}}^2-M_{\chi_1^0}^2} 
[2 C_1 C_2 -2 C_1 C_3], 
\end{equation}
where we have defined the following:
\begin{eqnarray}
C_1 & = & \frac{g_2 m_s N_{13}}{2 M_W \cos \beta}, \nonumber \\
C_2 & = & eQ y_1  +\frac{g_2}{\cos \theta_W} y_2 [T_3 -Q\sin^2 
\theta_W], 
\nonumber \\
C_3 & = & eQ y_1  -\frac{g_2}{\cos \theta_W} y_2 Q\sin^2 \theta_W.
\end{eqnarray}
Note that $C_1$ represents the coupling of the down type quark to the 
Higgsino portion of the neutralino.  $C_2$ and $C_3$ represent the 
couplings of bino to the left and right handed quark, respectively.  
Here, $T_3$ is the $SU(2)$ quantum number of the squark in question, $Q$ 
is the charge, $y_1$ denotes the photino fraction of the neutralino, 
while $y_2$ denotes the zino fraction.  They are given by:
\begin{eqnarray}
y_1&=& N_{11} \cos \theta_W + N_{12} \sin \theta_W, \nonumber \\
y_2&=& -N_{11} \sin \theta_W + N_{12} \cos \theta_W .
\end{eqnarray}

After approximating $y_2 \approx -\sin\theta_W$ and using 
$\tan\theta_W = g'/g$, a brief and straight-forward calculation 
yields a simple expression for the 
amplitude due to the exchange of the strange squarks: 
\begin{equation}
\label{squarkfinal}
B_{s}=\frac{-g_2 g'}{8M_{W} \cos \beta} 
\frac{N_{13}}{M_{\tilde{s}}^2- 
M_{\chi}^2}.
\end{equation}
Furthermore, we can write the masses $M_{\tilde{s}}$ and $ M_{\chi}$ 
in terms of the input parameters of mSUGRA. This is because the Yukawa 
couplings can be neglected in the RGEs.  Following the methods 
described in Ref.~\cite{Kazakov}, and using the Eq.~(\ref{m1mhalf}) we 
can write
\begin{equation}
M^2_{\tilde{s}_{R}}-M_{\chi}^2 \approx m_0^2+ 5.8 {m_{1/2}^2}.
\end{equation}

Using Eqs.~(\ref{adeq}) and (\ref{squarkfinal}), we can compare the 
squark exchange to the light Higgs boson exchange:
\begin{equation}
\frac{A^{d}}{B_{s}}=
\frac{2 \sin \alpha_H (N_{14} \cos \alpha_H + N_{13} 
\sin \alpha_H )( m_0^2+ 5.8 {m_{1/2}^2})}{m_{h}^2 N_{13}}.
\end{equation}
Here we have only kept the contribution of the strange quark to the 
Higgs exchange amplitude ($A^{d}$) as well.  Note that in general, the light 
Higgs boson contribution will dominate.  As expected, this is 
basically due to the fact that squarks are in general heavier than the 
lightest Higgs boson.  The squark exchange can be important only if 
the contribution from the exchange of the Higgs bosons is fine-tuned to be very small.

\section{General MSSM Framework}
\subsection{Definition of the Framework}
In this framework we relax our assumptions.  We keep the unification 
of the gaugino masses, but we drop the requirements that the scalar 
masses and the trilinear scalar couplings unify.  In addition, we drop 
the REWSB requirement (i.e., we take $m_{H_{1,2}}^{2}$ as 
independent parameters from $m_{0}$).  We assume that all scalar mass 
parameters at the {\it weak scale}\/ are equal: $m_{sq}$. This 
assumption is made in order to simplify the calculation, and it should 
not affect the general flavor of our results.  Of all trilinear 
couplings, we keep only $A_t$ and $A_b$ and we set all others to zero.  
Then, the free parameters are $\mu, M_2, \tan\beta, m_A, m_{sq}, A_t, 
A_b$.  We also relax the naturalness assumption, allowing the free 
parameters to have very large values.  Besides the relatively uniform 
scans of the parameter space, we also performed special scans in order 
to investigate the different conditions mentioned in the previous 
section.  The free parameter space is then:
\begin{eqnarray}
-300\,{\rm TeV} < \mu < 300 \, {\rm TeV},&& \; 0 < M_2 < 
300 \, 
{\rm TeV},\nonumber \\
95\,{\rm GeV} < m_A < 10 \, {\rm TeV},&& \; 200 \, {\rm GeV} < m_{sq} < 
50 \, 
{\rm TeV}, \nonumber \\
-3 < \frac{A_{t,b}}{m_{sq}} < 3 ,&& \; 1.8 < \tan \beta< 100. 
\label{mssmparamspace}
\end{eqnarray}

Again, a few comments are in order.  First, in this framework we 
observe higgsino-like (as well as bino-like) lightest neutralino.  In 
agreement with the Ref.~\cite{EFGOS}, we find very few light 
higgsino-like models, which will probably be explored soon by 
accelerator experiments.  Most of the higgsino-like models (with 
gaugino content $z_g < 0.01$) have  $M_{\chi}>450$ GeV, implying very 
large values of $m_{1/2}$. In particular, in higgsino like models 
$M_1 > \mu \approx M_{\chi}$; our results give $M_1 > 700$ GeV or, 
equivalently, $m_{1/2} > 1700$ GeV, which can be 
considered unnatural.  For these reasons, we choose not to analyze the 
higgsino case.  Second, $b \rightarrow s+\gamma$ is less constraining, 
but our results are still 
consistent with Refs.~\cite{bsgamma}, \cite{nohiggsino}.

\begin{figure}
[btp]
\includegraphics[width=\columnwidth]{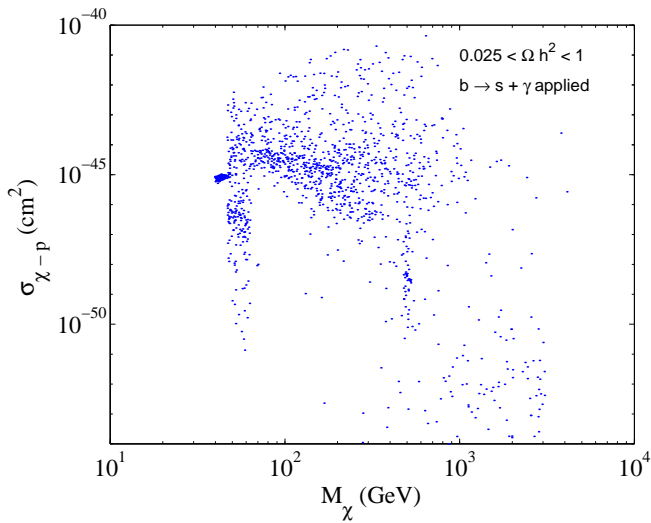}
\caption{The cross section for spin-independent $\chi$-proton 
scattering in the general MSSM framework is 
shown. A relatively conservative relic density cut is applied, 
$0.025 < \Omega h^{2} < 1$.   Note also that the constraint on the gaugino 
fraction, $z_g >10$, is applied.  Current accelerator bounds, 
including $b \rightarrow s + \gamma$ are imposed through the DarkSUSY code.}
\label{mssmcsvsmchi}
\end{figure}

We present the plot of $\sigma_{\chi-p}$ versus $M_{\chi}$ in this framework 
(Figure~\ref{mssmcsvsmchi}).  We do not pretend that 
Figure~\ref{mssmcsvsmchi} reflects all points accessible in a general 
MSSM. However, it does serve to show some generic differences from the 
mSUGRA case.  Namely, we can obtain much larger values for the 
neutralino mass because of the size of the parameter space.  In 
addition, the lower bound on $\sigma_{\chi-p}$ is also much lower than in the 
mSUGRA case.  We discuss the specifics of this below.

\subsection{Results and Analysis}
We concentrate only on the bino-like lightest neutralino.  All results 
in this section are presented with this assumption in mind.  In 
particular, we 
demand
$z_g = (N_{11}^2 + N_{12}^2)/(N_{13}^2 + N_{14}^2) > 10$. In this 
case, we can rely on the same approximations we used in the previous 
section.  In particular, the expansion of Eq.~(\ref{expansion}) is 
valid.  So, we can 
revisit the 5 different cases explored in the previous section.

\begin{figure}[hbtp]
\includegraphics[width=\columnwidth]{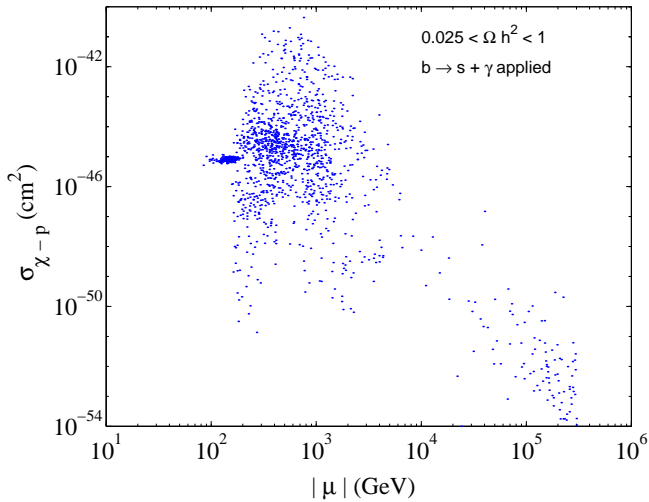}
\caption{The cross section for spin-independent $\chi$-proton 
scattering in the general MSSM framework is shown.  A relatively 
conservative relic density cut is applied, $0.025 < \Omega h^{2} < 1$.  
Note also that the constraint on the gaugino fraction, $z_g >10$, is 
applied.  Current accelerator bounds, including $b \rightarrow s + 
\gamma$ are imposed through the DarkSUSY code.}
\label{mssmcsvsmu}
\end{figure}

\begin{itemize}
\item Case 1: $N_{12} - N_{11}\tan\theta_W = 0$ cannot be satisfied 
because, as in the mSUGRA case, it implies $M_{\chi}<40$ GeV, which is 
ruled out by the relic density cut and the experimental limits (as 
shown on Figure~\ref{mssmcsvsmchi}).  The models that get close to 
satisfying this condition have very low contributions due to the Higgs 
bosons exchange, so this is one of the rare situations where the 
squark exchange is important.  
\item Case 2: $N_{14} \sin\alpha_H - 
N_{13} \cos\alpha_H = 0$ is now possible to satisfy because $\mu$ and 
$M_1$ are not related.  We indeed observe that the heavy Higgs boson 
exchange contribution is very small in this case, but since the light 
Higgs boson exchange dominates, $\sigma_{\chi-p}$ is kept relatively high in 
value.  
\item Case 3: $N_{14} \cos\alpha_H + N_{13} \sin\alpha_H = 0$ 
is also possible to satisfy.  As in the mSUGRA case, the light Higgs 
boson exchange is small and heavy Higgs boson exchange dominates.  
Unlike in the mSUGRA case, the accelerator bounds do not rule out
these models.  
\item Case 4: Since the naturalness constraint 
has been relaxed, $\mu$ is allowed to have very large values.  Then 
$N_{12}$, $N_{13}$, and $N_{14}$ can be driven small, which in turn 
would make the Higgs boson exchange contribution small.  Intuitively, 
large $|\mu |$ implies that the neutralino is a very pure bino, for 
which the Higgs boson scattering channels vanish.  If the squark 
masses are kept large as well, the squark contribution will be small 
too, making the total elastic scattering cross section very small.  
This is illustrated in Figure~\ref{mssmcsvsmu} - the lowest values of 
$\sigma_{\chi-p}$ are obtained for the largest values of $|\mu |$.
\item Case 5: As discussed in the mSUGRA case, if 
($\frac{M_1}{\mu}\tan\beta$) is negative and sufficiently large, 
$N_{14}$ can change sign.  Following 
through Eqs.~(\ref{fhH}) and (\ref{adeq}), this effect can induce 
destructive interference between the $A^u$ and $A^d$ terms in 
Eq.~(\ref{sigma}) and cause the overall $\sigma_{\chi-p}$ to vanish.
We observed this cancellation in the mSUGRA case for $M_{\chi} > 120$ GeV,
and we also observe it in the general MSSM case. 
In Table 1 we present some of the models in which this kind of 
cancellation takes place. Besides the models at large $M_{\chi}$, we
also observe models with relatively low $M_{\chi} (\approx M_1)$ and very low 
values of $\sigma_{\chi-p}$ (see rows 3 and 4 of Table 1). These models
were not allowed in the mSUGRA framework due to the $M_1 - \mu$ relationship
(determined by the RGE's and the REWSB assumption), 
which does not exist in the general MSSM framework.
\end{itemize}

\begin{table*}[!ht]
\centering
\begin{tabular}{|r|r|r|r|r|r|r|r|}
\hline
$\mu$ (GeV) & $M_1$ (GeV) & $m_{sq}$ (GeV) & $m_A$ (GeV) & $\tan\beta$ 
& $A_t / m_{sq}$ & $A_b / m_{sq}$ & $\sigma_{\chi-p} \; ({\rm cm^2})$ \\
\hline
-1794 & 502 & 3792 & 1004 & 10.1 & 1.2 & 2.5 & 7.9$\times 10^{-51}$ \\
\hline
-2109 & 534 & 3211 & 1087 & 11.8 & -1.3 & 1.0 & 8.4$\times 10^{-51}$ \\
\hline
-195 & 55 & 2995 & 1120 & 10.2 & -2.0 & 2.3 & 2.1$\times 10^{-50}$ \\
\hline
-182 & 61 & 2891 & 1099 & 7.0 & -0.6 & -0.1 & 1.6$\times 10^{-50}$ \\
\hline
-274 & 163 & 325 & 1944 & 3.8 & 0.6 & 2.5 & 7.8$\times 10^{-50}$ \\
\hline
\end{tabular}
\caption{Some of the models in the general MSSM framework with very 
low values of $\sigma_{\chi - p}$.}
\end{table*}

In other words, in the general MSSM framework it is possible to obtain
low values of $\sigma_{\chi-p}$ in several different ways: either by
tuning parameters to suppress the contribution of the heavy or the light 
Higgs boson exchange, 
or by allowing parameters (such as $\mu$) to be very large, which suppreses 
both the heavy and the light Higgs boson exchange channels, or by 
fine-tunning parameters to achieve a complete cancellation of
terms in the Eq. (\ref{sigma}). As a result, the lower bound on 
$\sigma_{\chi-p}$ vanishes and it is beyond reach of the present
and the proposed direct detection experiments.

\section{Conclusion}

We summarize our results as follows. The main contributions to the 
cross section for spin independent elastic scattering of neutralinos
off nucleons come from the exchange of the Higgs bosons and squarks.
The contribution of the squark exchange is usually much smaller, 
being of 
importance only when Higgs boson exchange contribution is very small.

We investigate different conditions which could lead to small Higgs 
boson exchange contribution.  We find that in mSUGRA framework, with 
the free parameter ranges defined in Eq.~(\ref{paramspace}), these 
conditions are not satisfied due to the relationship between 
parameters $M_1$ and $\mu$ (coming from the unification and radiative 
electroweak symmetry breaking assumptions), the naturalness assumption 
(which keeps different parameters from becoming very large) and the 
accelerator constraints. We find that the light Higgs boson exchange 
dominates over 
the other channels and it leads to $\sigma_{\chi-p} > 10^{-46} {\rm cm^2}$.  
Equivalently, this yields an event rate $> 0.1 {\rm \; ton^{-1} 
day^{-1}}$ in $^{73} {\rm Ge}$ target, which could be within reach of 
the future direct detection experiments.  However, if we 
expand the mSUGRA framework to allow larger neutralino masses, the
lower bound on $\sigma_{\chi-p}$ vanishes for those large masses due to 
the occasional complete cancellation of terms in the Eq. (\ref{sigma}). 

In the more general MSSM models (as defined in and above 
Eq.~(\ref{mssmparamspace})), the situation is significantly different.  
The light and/or heavy Higgs boson exchange channel can now be suppressed
either by tuning parameters to satisfy Cases 2) or 3), or by allowing
the free parameters (such as $\mu$) to be very large. Moreover, the
complete cancellation of terms in the Eq. (\ref{sigma}) is now
possible even at low values of $M_{\chi}$ because the relationship 
between $M_1$ and $\mu$ that existed in the mSUGRA framework 
due to radiative electroweak symmetry breaking, is relaxed in the general
MSSM framework. As a result, the lower bound on $\sigma_{\chi-p}$ is much
lower than in the mSUGRA framework and it is beyond reach of the current or 
proposed direct detection experiments.

\section{Acknowledgements}
VM thanks Bernard Sadoulet and Richard Gaitskell for discussions and
suggestions regarding this work. PG thanks Bernard Sadoulet for hospitality 
at the CfPA.
The work of HM and AP was supported in part by 
the Director, Office of Science, Office of High Energy and Nuclear Physics, 
Division of High Energy Physics of the U.S. Department of Energy under 
Contract DE-AC03-76SF00098 and in part by the National Science Foundation 
under grant PHY-95-14797.  AP is also supported by a National Science 
Foundation Graduate Fellowship.
The work of VM was supported by the Center for Particle Astrophysics, 
a NSF Science and Technology Center operated by the University of California, 
Berkeley, under Cooperative Agreement No. AST-91-20005 and by the 
National Science Foundation under Grant No. AST-9978911.

\end{document}